\documentclass[manuscript]{aastex}

\usepackage{natbib, aas_macros}
\usepackage{amsmath}
\usepackage{color}

\usepackage{ulem}

\definecolor{MyDarkBlue}{rgb}{0,0.08,0.45}
\definecolor{MyDarkRed}{rgb}{0.8,0.1,0.08}
\definecolor{Red}{rgb}{1.0,0.0,0.2}
\definecolor{Blue}{rgb}{0,0.08,0.95}
\definecolor{LightGrey}{rgb}{0.7,0.7,0.7}

\begin{document}
\title{Solar overshoot region and small-scale dynamo with realistic
energy flux}
\author{H. Hotta}
\affil{Department of Physics, Graduate School of Science, Chiba
university, 1-33 Yayoi-cho, Inage-ku, Chiba, 263-8522, Japan
}
\email{ hotta@chiba-u.jp}
 \begin{abstract}
  We carry out high resolution calculations of the solar overshoot
  region with unprecedentedly realistic parameters, especially the
  small energy flux compared with $\rho c_\mathrm{s}^3$, where $\rho$
  and $c_\mathrm{s}$ are density and speed of sound. Our main purpose
  is to investigate behavior of the overshoot and the small-scale dynamo with
  parameters as close as possible to thoes of the Sun.
  Our calculations show that the bottom part of the convection zone becomes
  subadiabatic, which efficiently suppresses downflows. As a result
  we see a steep transition from the CZ to the RZ whose width is
  estimated 0.4\% of the local pressure scale height.
  This result is consistent with a semi-analytic
  convection/overshoot model. We also find that the small-scale dynamo
  becomes efficient with a smaller energy flux. 
  The sudden suppression of the downflows around
  the base of the convection zone increases the efficiency
  of the small-scale dynamo.
 \end{abstract}

\keywords{Sun: interior --- Sun: dynamo --- Stars: interiors}
\clearpage
\section{Introduction}
The Sun is composed of a convectively stable radiative zone (RZ: 
inner 70\% of radius) and an unstable convection zone (CZ: outer 30\%).
Between these two regions, there is thought to be overshoot region
(OS). Although a proper description of the OS is crucial to understand
the solar rotation profile, especially the tachocline, and the global solar dynamo
\citep[e.g.][]{1993ApJ...408..707P,2011ApJ...742...79B}, there has been
a substantial discrepancy between its observations, and what is expected based on
analytical approachs, and numerical calculations.

On the observational side, helioseismology uses
acoustic mode, i.e., the sound wave. The discontinuity of the temperature
gradient introduces a characteristic oscillatory component in the
frequencies. Thus the helioseismology essentially estimates the temperature
gradient to see the structure of the OS. Since the CZ is slightly
superadiabatic and the RZ is strongly subadiabatic, this transition
can be detected.
\cite{1997MNRAS.288..572B} sets an upper limit on the thickness of the
OS ($d_\mathrm{os}$) of 0.05$H_p$,
where $H_p\sim 6\times10^9\ \mathrm{cm}$ is the local pressure scale
height at the base of the CZ. Other studies using
helioseismology also show similar thin OS
\citep{1994A&A...283..247M,1994MNRAS.268..880R,1994MNRAS.267..209B,2016LRSP...13....2B}.

On the analytic side, early simplified models of the OS show results of the rather thick OS of
$d_\mathrm{os}=(0.2-0.4)H_p$
\citep{1982A&A...113...99V,1984ApJ...282..316S,1986A&A...157..338P}.
These three models used different assumptions,
i.e., a non-local mixing length model, overturning convective rolls, and
a model of the downflow plumes, respectively.
The models produced consistent result, independently of the different assumptions adopted.

On the numerical side, a lot of numerical calculations have been performed. Two- and
three-dimensional numerical calculations show thicker OS of $d_\mathrm{os}=(0.4-2)H_p$
\citep{1993A&A...277...93R,1994ApJ...421..245H,1995A&A...295..703S,1998A&A...340..178S,2002ApJ...570..825B}.
\cite{2004ApJ...607.1046R} points out that the discrepancy between the
simplified and numerical models is caused by the difference of the
imposed energy flux from the bottom of the calculations -- here the
semi-analytic model of \cite{2004ApJ...607.1046R}
also shows the thickness of the OS of 0.4$H_p$ with a sharp transition
region to the RZ. The reason for the high energy flux in the simulations is
in order to avoid the severe constraint by the CFL condition when the correct
flux is used, however the large energy flux increases the
convection velocity and the overshoot penetration depth
(see the detailed explanation in \S \ref{difficulty}).

\par
Recent high resolution calculations show that the base of the CZ
is important for both the small- and large-scale dynamo
\citep{2015ApJ...803...42H,2016Sci....351..1427}. \cite{2015ApJ...803...42H}
show that small-scale magnetic field reaches the equipartition level of
the turbulent flow in the bulk of the CZ, when the resolution is high
enough to resolve turbulent inertia scale well.
Especially at the base of the CZ, the strong magnetic field
suppresses the small-scale flow and the turbulent RMS velocity is
decreased by a factor of two. The magnetic energy there exceeds 90\% of the
kinetic energy.
\cite{2016Sci....351..1427} find that the strong
small-scale magnetic field at the base of CZ has a significant effect on
the large-scale dynamo. The strong small-scale magnetic field suppresses the small-scale
flows which otherwise tends to destroy large-scale features. Thus this suppression
of the small-scale flow allows large-scale magnetic field to be generated even
with large Reynolds numbers. Both of the above papers, however, investigate only the
CZ, i.e., the RZ is excluded. Thus the overshoot region is explicitly ignored in these
two papers. We need to improve our understanding of the OS for both the small- and large-scale dynamos.

Other three-dimensional models including the RZ argue that the OS is an important place
for the large-scale dynamo in terms of the storage of the large-scale
magnetic flux \citep{2006ApJ...648L.157B,2013ApJ...778...11M,2016ApJ...819..104G}. These
studies typically adopt large energy flux or
softer RZ (smaller absolute value of the superadiabaticity) to avoid
numerical difficulties discussed in the next subsection.
The values in typical calculations are shown in the next section.
These studies thus likely overestimate the thickness of the OS and its
ability to store large-scale magnetic flux.
In this paper, we address this issue by adopting more realistic solar parameters,
such as the imposed energy flux.

\subsection{The difficulty of including a realistic radiation zone}\label{difficulty}
In order to discuss the thermal convective stability, a non-dimensional
number, superadiabaticity, is useful. The superadiabaticity
($\delta$) is defined as:
\begin{eqnarray}
 \delta = \nabla - \nabla_\mathrm{ad} =
  -\frac{H_p}{c_\mathrm{p}}\frac{ds}{dr},
\end{eqnarray}
where $\nabla=d\ln T/d\ln p$ is the temperature gradient and
$\nabla_\mathrm{ad}$ is the adiabatic one. $r$, $T$, $p$, $s$,
$c_\mathrm{p}$, and $H_p=(d\ln p/dr)^{-1}$ are the local radius, 
temperature, pressure, specific entropy,  heat capacity at
constant pressure, and  pressure scale height, respectively.
The estimated values in the solar CZ and RZ are
$\delta\sim10^{-7}$ and $-10^{-1}$, respectively \citep{1996Sci...272.1286C}.
The superadiabaticity in the CZ is related to the ratio of the
convection velocity ($v_\mathrm{c}$) and the sound speed ($c_\mathrm{s}$), i.e.,
$\delta\sim(v_\mathrm{c}/c_\mathrm{s})^2$ using mixing length theory
\citep{1953ZA.....32..135V}. Thus the small superadiabaticity implies
that the sound speed is much faster than the
convective velocities. This severely constrains the CFL condition and requires a very
small time step in terms of the thermal convection is required. In recent numerical
calculations, this difficulty is avoided either by using the anelastic approximation
\citep[e.g.][]{2000ApJ...532..593M} or the reduced speed of sound technique
\citep[RSST:][]{2012A&A...539A..30H,2015ApJ...798...51H}
where the propagation of the sound wave is changed.

The small superadiabaticity in the solar CZ is the consequence of the relatively small imposed energy flux.
The energy flux in the CZ can be roughly estimated as $F_{\odot}\sim \rho v_\mathrm{c}^3$, where $\rho$ is
the density. Thus, the superadiabaticity can be expressed as a normalized flux 
$\delta \sim \left({v_\mathrm{c}}/{c_\mathrm{s}}\right)^2\sim \bar{F}_\odot^{2/3} $,
where the normalized energy flux is expressed with $\bar{F}_\odot=F_\odot/(\rho c_\mathrm{s}^3)$, which is estimated
 to be $\bar{F}_\odot\sim5\times10^{-11}$ at the base of the solar CZ with Model S
 \citep{1996Sci...272.1286C}.
\par
In the RZ, internal gravity waves oscillate
with the Brunt-V\"{a}is\"{a}l\"{a} frequency. The
Brunt-V\"{a}is\"{a}l\"{a} frequency ($N$) is 
\begin{eqnarray}
 N^2 = \frac{g}{c_\mathrm{p}}\frac{ds}{dr} = -\frac{1}{\gamma}\frac{c_\mathrm{s}^2}{H_p^2}\delta,
\end{eqnarray}
where $g$ is the gravitational acceleration.
The square root of the absolute value of the superadiabaticity in the RZ is
the ratio of the time
scales of the internal gravity wave ($1/N$) and the sound wave
($H_p/c_\mathrm{s}$). Since the square root of the superadiabaticity
in the solar RZ is close to unity, the time scale for the
internal gravity wave and the sound wave is similar in the solar RZ.
When the vertical wave length is close to
the pressure scale height $(\sim H_p)$, the phase velocity of the
internal gravity wave ($N \lambda_\mathrm{v}/(2\pi)\sim
c_\mathrm{s}\sqrt{-\delta/\gamma}$, where $\lambda_\mathrm{v}$ is the
vertical wave length) can be comparable to the speed of sound.
Thus internal gravity waves also require a time step which is very small
in terms of the thermal convection. This cannot be avoided with either the anelastic
approximation or the RSST.
\par
The essential difficulties to numerically investigate the OS are
summarized as:
\begin{enumerate}
 \item The speed of sound is much faster than the convection velocity in
       the  CZ due to the small energy flux
       ($\overline{F}_\odot\sim5\times10^{-11}$).       
 \item The internal gravity wave has a similar time scale
       (Brunt-V\"{a}is\"{a}l\"{a} frequency) to the sound wave in
       the RZ.
\end{enumerate}

Additional numerical complications arise because of 
the significant difference of the superadiabaticities in the CZ
and RZ, which leads to the expectation that the OS or the transition region to the
RZ should be very thin. This requires the use of a large number of grid points to resolve this
thin layer.
When the anelastic approximation is used, the numerical calculation
is likely to use softer RZ. For example, \cite{2000ApJ...532..593M}
adopt $\delta\sim-10^{-4}$ in the RZ (see Fig. 1 of
\cite{2000ApJ...532..593M}). 
Most compressible calculations are done using a very large
energy flux of $\overline{F}_\odot\sim10^{-3}$ to $10^{-4}$
\citep{1994ApJ...421..245H,2002ApJ...570..825B,2010AN....331...73K}.
This likely overestimates the thickness of the OS.
\par

The purposes of the present study are to:
\begin{enumerate}
 \item Investigate the dependence of the overshoot depth on the
 energy flux. Although a calculation with the real solar value
 ($\bar{F}_\odot\sim5\times10^{-11}$) is still difficult to carry out,
       we achieve unprecedentedly small energy flux
       $\overline{F}_0=5\times10^{-7}$.
       We find a scaling law for the normalized energy flux and overshoot
       depth and estimate a real solar overshoot depth.
 \item Fill the gap between the observations and the simplified
       models. Our calculations includes almost all physics with
       most realistic parameters of the CZ, the RZ, and the OS. This possibly
       gives an insight
       into the realistic solar overshoot region.
 \item Compare overshoot depths with and without the magnetic field. Our
       previous calculation revealed that the small-scale dynamo is very
       efficient in the solar CZ \citep{2015ApJ...803...42H}. The
       magnetic field in high-resolution calculations possibly
       influences the behavior of the overshoot layer.
 \item Investigate the dependence of the small-scale dynamo on the
       imposed energy flux, i.e., the superadiabaticity in the CZ.
 \item Explore appropriate boundary conditions for the magnetic
       field at the base of the CZ. Although currently the
       anelastic approximation and the RSST are not       
       useful for investigations of the OS, these method are
       necessary in order to cover the time scale of large-scale dynamo. We need to
       confirm appropriate boundary condition for the important place,
       the base of the CZ.
\end{enumerate}

 \section{Model}
We solve the three-dimensional magnetohydrodynamic (MHD) equations  in a local
Cartesian geometry $(x,y,z)$. We define the $x$-direction as that of
gravity. The  $y$- and $z$-directions are horizontal directions. The MHD
equations with the gravity and radiative heating and cooling are:
\begin{eqnarray}
 \frac{\partial \rho}{\partial t} &=& -\nabla\cdot(\rho{\bf v}),\\
 \frac{\partial}{\partial t}(\rho{\bf v}) &=& -\nabla\cdot
  \left[
   \rho{\bf vv} + \left(p{\bf I} + \frac{B^2}{8\pi}{\bf I}-\frac{\bf BB}{4\pi}
		  \right) 
				  \right]- \rho g{\bf e}_x,\\
 \frac{\partial {\bf B}}{\partial t} &=& \nabla\times({\bf v}\times{\bf
  B}),\\
 \frac{\partial E_\mathrm{total}}{\partial t} &=& -\nabla\cdot
  \left[\left(
	 E_\mathrm{total} + p + \frac{B^2}{8\pi}
		      \right){\bf v}
  -\frac{({\bf v}\cdot{\bf B}){\bf B}}{4\pi} - \kappa_\mathrm{r}\nabla T
				 \right]-\rho v_xg + \Gamma,\\
 E_\mathrm{total} &=& \frac{p}{\gamma-1} + \frac{1}{2}\rho v^2 +
  \frac{B^2}{8\pi},
\end{eqnarray}
where ${\bf v}$, ${\bf B}$, $E_\mathrm{total}$, $\kappa_\mathrm{r}$, and
$\Gamma$ are the fluid velocity, the magnetic field, the total energy,
the radiative diffusivity, and the time independent cooling around the
top boundary, respectively.
We adopt the ideal equation of state $p=R\rho T$, where $R$ is the gas
constant and the ratio of the
heat constant is $\gamma=c_\mathrm{p}/c_\mathrm{v}=5/3$, where
$c_\mathrm{v}$ is the heat capacity at constant volume.
We use forth-order centered differences to discretize the problem in space, and the
four-step Runge-Kutta method for time the integration \citep{2005A&A...429..335V}.
The artificial viscosity suggested by
\cite{2009ApJ...691..640R} and \cite{2014ApJ...789..132R} is used to maintain stability.
The artificial viscosity adopts slope-limiter which
defines the slope of variables in a grid. Thanks to the slope limiter,
we are able to minimize the diffusion in a smooth profile and maximize
it at a steep gradient.

For initial conditions, we take a  hydrostatic equilibrium with
 $dp/dx =-\rho g$. The temperature gradient is expressed as
 $\nabla=d\ln T/d\ln p$. Thus the initial conditions are
\begin{eqnarray}
 T_0 &=&T_\mathrm{b}\left[1-\nabla\frac{x}{H_\mathrm{b}}\right],\\
 p_0 &=&p_\mathrm{b}\left[1-\nabla\frac{x}{H_\mathrm{b}}\right]^{1/\nabla},\\
 \rho_0 &=&\rho_\mathrm{b}\left[1-\nabla\frac{x}{H_\mathrm{b}}\right]^{1/\nabla-1},
\end{eqnarray}
where $T_\mathrm{b}$, $p_\mathrm{b}$, $\rho_\mathrm{b}$, and
$H_\mathrm{b}=-(d\ln p/dx)^{-1}|_{x=0}=p_\mathrm{b}/\rho_\mathrm{b} g$ are
the temperature, the pressure, the density, and the pressure scale
height at $x=0$, which is the base of the CZ.
The calculation domain is $(-1,0,0)<(x,y,z)/H_\mathrm{b}<(1.8,5.6,5.6)$ for
cases with the RZ and
$(0,0,0)<(x,y,z)/H_\mathrm{b}<(1.8,5.6,5.6)$ for cases without the
RZ.
Although calculations in this study are non-dimensional,
the horizontal box size
corresponds to $340\ \mathrm{Mm}\sim0.5R_\odot$ when we fix the spatial
scales by reference to $H_\mathrm{b}=60\ \mathrm{Mm}$.

We use a stable stratification below $x=0$ (Layer I) and
initially neutral above $x=0$ (Layer II). In the Layer I, the superadiabaticity
$\delta$ is set to -0.2, where the adiabatic temperature gradient is adopted in the Layer I
(in the initial condition).
We adopt an impenetratable boundary condition for the velocity, i.e.
$v_x=\partial v_y/\partial x=\partial v_z/\partial x=0$ at both the top
and bottom boundaries. At the top boundary only the vertical magnetic
field is allowed, $\partial B_x/\partial x=B_y=B_z=0$. Only horizontal
magnetic field is allowed at the bottom boundary,
 $B_x=\partial B_y/\partial x=\partial B_z/\partial x=0$.
 Periodic boundary condition is adopted for the all variables in
 the horizontal directions.
 
After calculations start, the Layer II becomes superadiabatic.
$\Gamma$ is the time independent cooling around the top boundary expressed as:
\begin{eqnarray}
 F &=&
  F_0\exp\left[-\left(\frac{x-x_\mathrm{max}}{d_\mathrm{max}}\right)^2\right],
  \label{cooling}
  \\
 \Gamma &=& -\frac{dF}{dx},
\end{eqnarray}
where $x_\mathrm{max}=1.8H_\mathrm{b}$ is the location of top boundary
and we set $d_\mathrm{max}=0.2H_\mathrm{b}$.
$F_0$ is the imposed flux at the base by the radiative diffusion.\par
\begin{figure}[htbp]
  \centering
  \includegraphics[width=16cm]{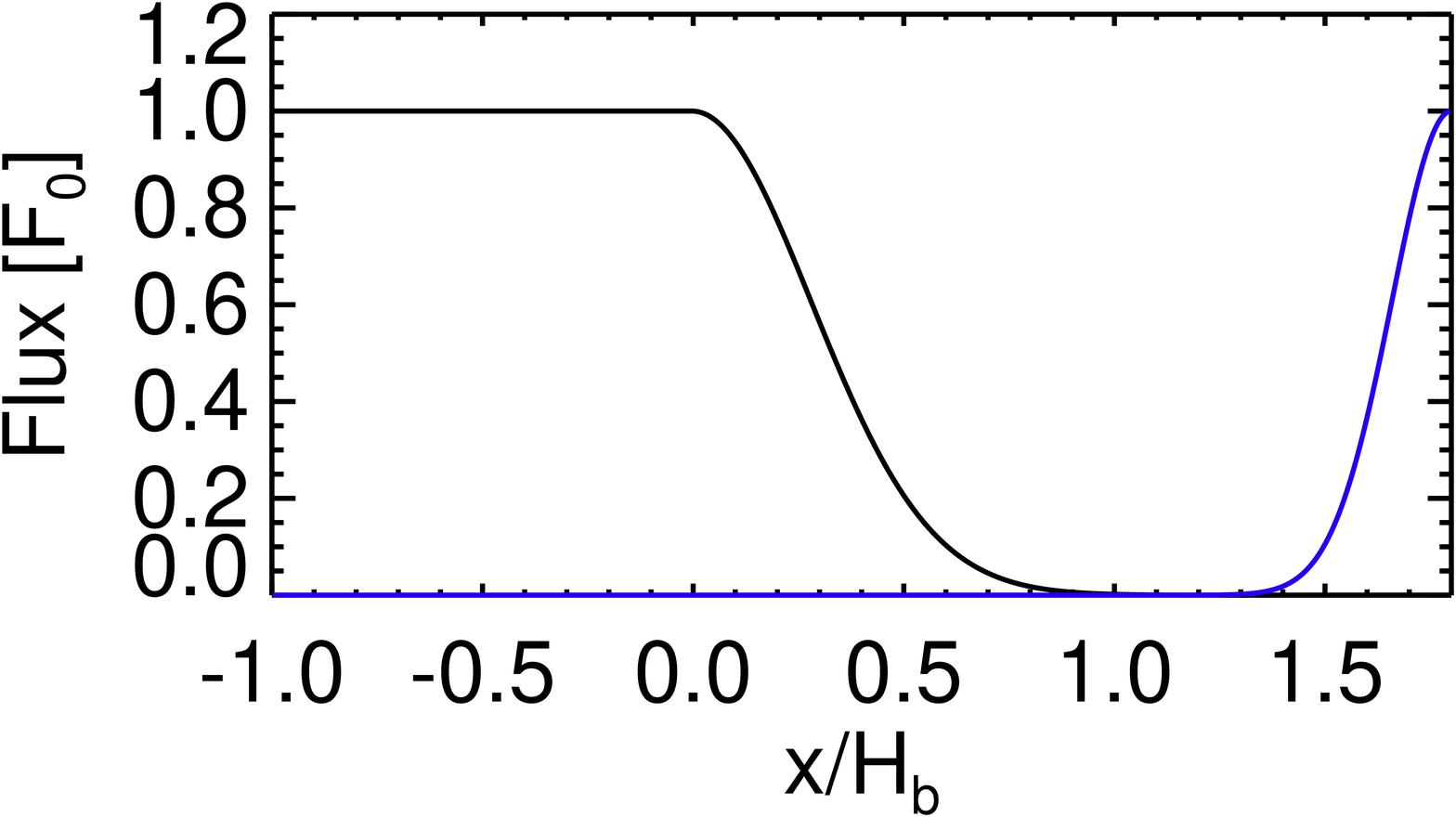}
  \caption{The radiative flux ($-\kappa_\mathrm{r}dT_0/dx$: black) and
  the cooling around the top boundary (F: blue) are shown. The fluxes
  are normalized with the imposed flux $F_0$.
  \label{rad_flux}}
 \end{figure}

Since all the energy flux is transported by the radiation in the RZ,
the radiative diffusion coefficient is defined only by the imposed energy
flux and the temperature gradient as:
\begin{eqnarray}
 \kappa_\mathrm{r} = -\frac{F_\mathrm{0}}{dT_0/dx}\ \ (x<0).
\end{eqnarray}
Around the base of the CZ, some fraction of the energy is transported
by the radiation in the Sun. Thus we mimic this situation with
decreasing the radiative diffusivity as:
\begin{eqnarray}
 \kappa_\mathrm{r} = -\frac{F_\mathrm{0}}{dT_0/dx}
  \exp
  \left[
  -\left(
   \frac{x}{d_\mathrm{min}}
	   \right)^2
  \right]
\ \ (x\ge0),
\end{eqnarray}
where $d_\mathrm{min}=0.4H_\mathrm{b}$.
Fig. \ref{rad_flux} shows the radiative flux
($-\kappa_\mathrm{r}dT_0/dx$: black
) and
  the cooling around the top boundary (F: blue).

\par
Although we start calculations with the simple analytic solution of 
hydrostatic balance, this initial condition changes in the RZ  in order to
reach an energy balance in the overshoot
layer.
We define normalized energy flux as
$\bar{F}_0 = F_0/\rho_\mathrm{b}c_\mathrm{b}^3$,
where $\rho_\mathrm{b}$ and $c^3_\mathrm{b}$ are the density and the
speed of sound at the base of CZ.

When the normalized energy flux is small ($\bar{F}_0<<1$), this change
takes a long time compared to thermal convection. In order
to shorten this process, to keep the calculation reasonable, we adopt
a similar method to \cite{2000ApJ...532..593M} and
\cite{2011ApJ...742...79B}. In these papers, they increased the radiative
diffusion in the RZ ($x<0$) in response to the enthalpy flux. In this paper,
we increase the radiative diffusion $\kappa_\mathrm{r}$ to:
\begin{eqnarray}
 \kappa_\mathrm{r} = \frac{F_0 - F_\mathrm{s}}{d\overline{T}/dt},
\end{eqnarray}
where the total ($F_\mathrm{s}$), the enthalpy ($F_\mathrm{e}$), the
kinetic ($F_\mathrm{k}$), and Poynting ($F_\mathrm{m}$) fluxes are
defined as:
\begin{eqnarray}
  F_\mathrm{s} &=& F_\mathrm{e} + F_\mathrm{k} + F_\mathrm{m} +
   F_\mathrm{a},\\
  F_\mathrm{e} &=& \int c_\mathrm{p}T(\rho v_x -
   \overline{\rho v_x})dS/\int dS,\\
 F_\mathrm{k} &=& \int \frac{1}{2}v^2(\rho v_x - \overline{\rho
  v_x})dS/\int dS,\\
 F_\mathrm{m} &=& \int \frac{1}{4\pi}(B^2v_x - {\bf v}\cdot{\bf
  B}B_x)dS
  /\int dS,
\end{eqnarray}
where overlines on the variables expresses horizontal averages.
Here $F_\mathrm{a}$ is the energy flux due to the artificial viscosity on the
internal energy \citep[see the detail in][]{2014ApJ...789..132R}. Here,
we investigate the downflow penetration and dynamo in this state
(further discussion is seen around Fig. \ref{flux}).
\par

In order to compare results from simulations with significantly different energy
fluxes, we define ``convection
 velocity ($v_\mathrm{c}$)''.
 As discussed in the introduction, the normalized energy
 flux determines
 the superadiabaticity in the CZ as
 $\delta\sim\bar{F}_0^{2/3}$.
 We change the normalized energy flux
 and study the dependence of the convection and dynamo behavior.
 From the relation of the energy flux and the convection velocity
 $F_0\sim\rho_0v_\mathrm{c}^3$, the convection velocity is
 expressed as:
 \begin{eqnarray}
  v_\mathrm{c} = \bar{F}_0^{1/3}c_\mathrm{b}.
 \end{eqnarray}
 We use this value ($v_\mathrm{c}$) for the normalization.\par
\begin{table}
\begin{center}
 \caption{ Free parameters for calculations.
 As a total, $3\times4\times2=24$ cases are calculated.
 The numbers in the square brackets show the number of grid points in x
 direction for the cases without the RZ.
 \label{table}}
\begin{tabular}{l|c}
 \hline
 \hline
 Grid points & \\
 ($N_x, N_y, N_z$) &
  ($128[88], 256, 256$), ($256[176], 512, 512$),
     ($512[376], 1024, 1024$)\\
 Normalized flux $\overline{F}_0$ &
     $5\times10^{-5}$, $5\times10^{-6}$, $5\times10^{-7}$ (w/ RZ),
     $5\times10^{-6}$ (w/o RZ)\\
 Magnetic field & Yes, No \\
\hline
\hline
\end{tabular}
\end{center}
\end{table}

We study 24 cases in total. All the cases are summarized in Tables
\ref{table}. For the cases with a RZ, we
choose three different normalized energy fluxes,
$\overline{F}_0=5\times10^{-5}$, $5\times10^{-6}$, and $5\times10^{-7}$.
In the highest resolution cases, the grid spacing is
$\Delta x = \Delta y = \Delta z\sim5.5\times10^{-3}H_\mathrm{b}$. This
corresponds to 330~km on the Sun.
 Thermal convection in the CZ is largely uninfluenced by
 imposed energy flux, when we consider quantities normalized with the 
 convection velocity
 $v_\mathrm{c}=\overline{F}_0^{1/3}c_\mathrm{b}$. Thus we do
 not change the normalized energy flux ($\overline{F}_0=5\times10^{-6}$) for the cases
 without the RZ.
 We carry out the calculations with three
 different resolutions, High, Medium, and Low. The numbers of the grid
 points are shown in Tables \ref{table}. In order to use identical grid
 spacings for the cases with and without RZ, the number of grid points in
 the $x$ direction is changed, correspondingly. We mainly analyze the case
 high resolution series.
 For all the cases, we carry
 out calculations with and without the magnetic field. For the cases with
 the magnetic field, we initially inject a weak seed magnetic field.
All graphics and analysis presented in this study correspond to the High
series (unless otherwise stated)
 \section{Result}

 \begin{figure}[htbp]
  \centering
  \includegraphics[width=12cm]{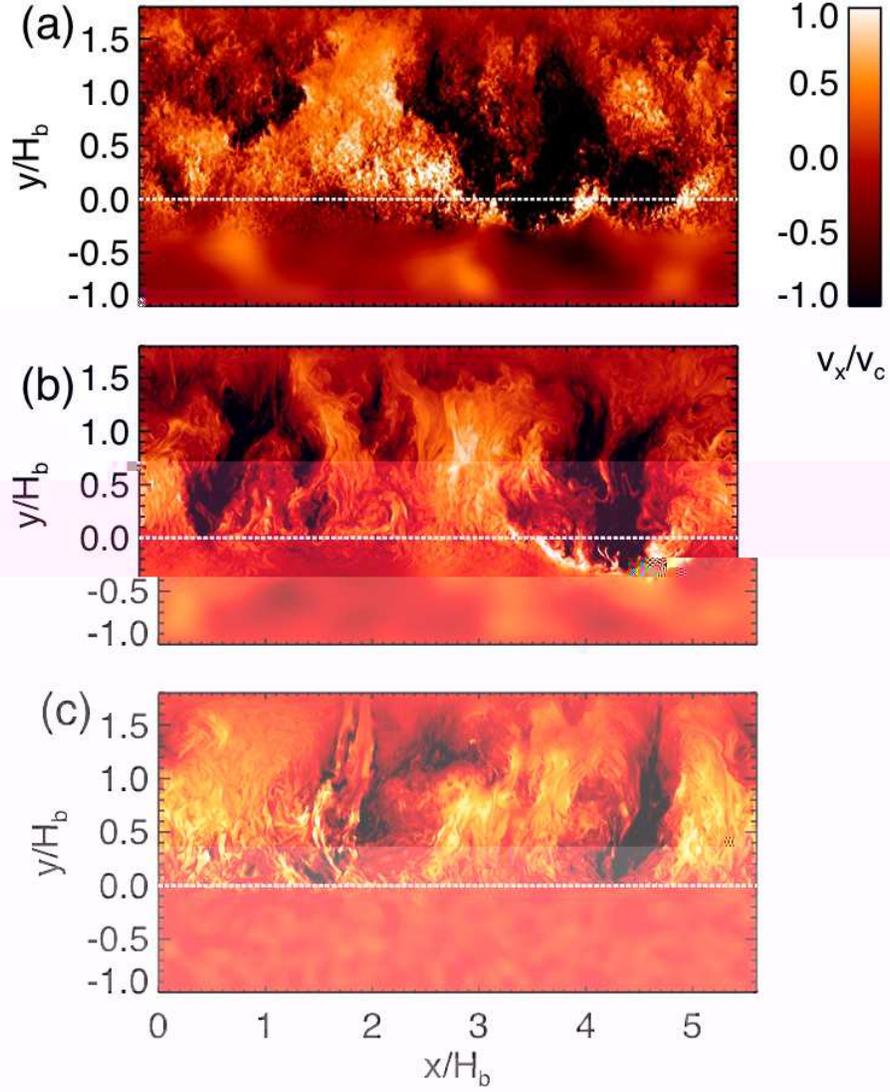}
  \caption{
  Vertical cross sections of the vertical velocity
  ($v_x$).
  The results with (a)
  $\overline{F}_0=5\times10^{-5}$ without the magnetic field,
  (b) $\overline{F}_0=5\times10^{-5}$ with the magnetic field, and
  (c) $\overline{F}_0=5\times10^{-7}$ with the magnetic field are shown.
  White dashed lines shows $x=0$,
  where the location of the transition between the CZ and the RZ.
  \label{ctvx}}
 \end{figure}

 \begin{figure}[htbp]
  \centering
  \includegraphics[width=16cm]{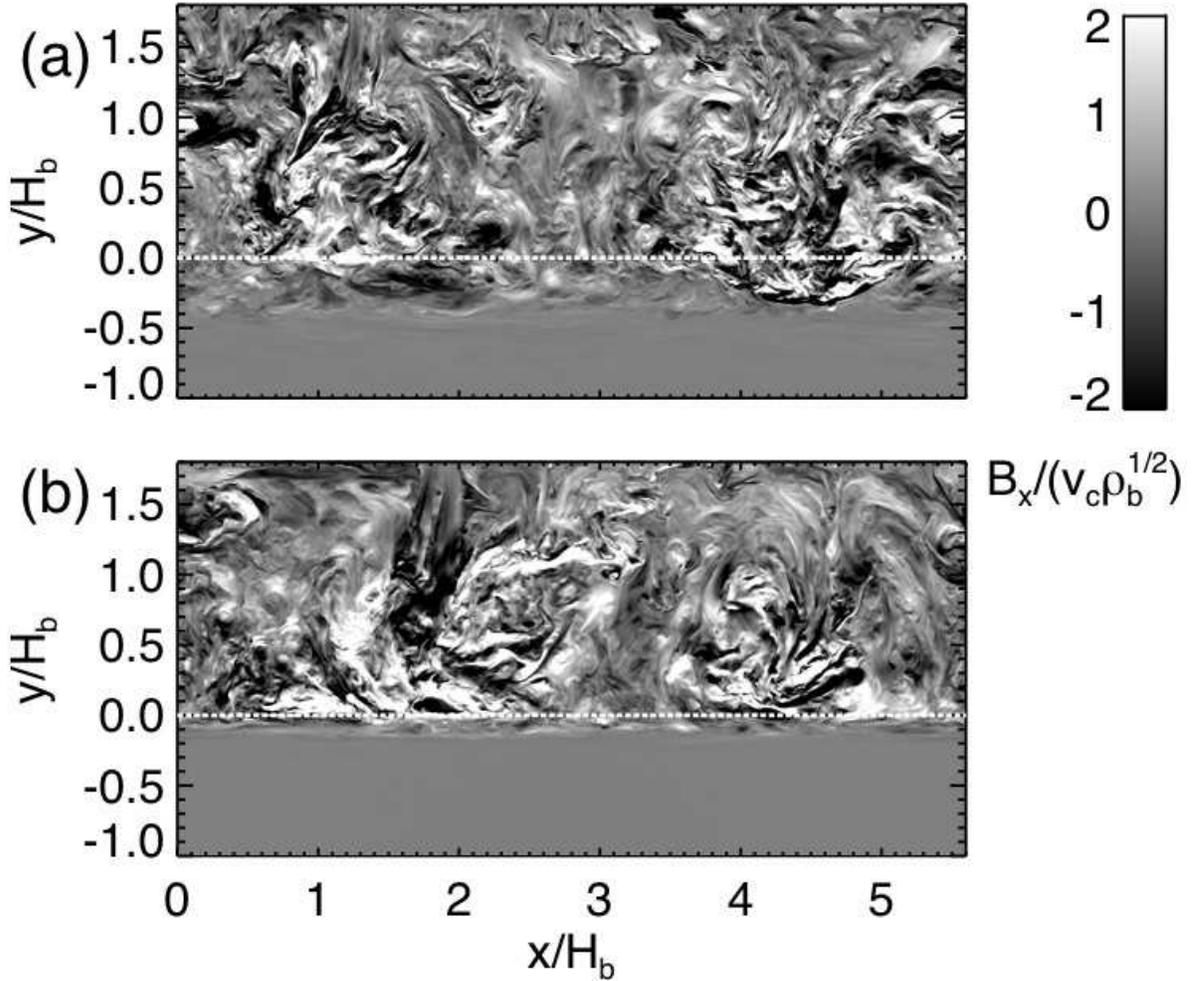}
  \caption{ 
  Vertical cross sections of the vertical magnetic field
  ($B_x$).
  The results with
  (a) $\overline{F}_0=5\times10^{-5}$ and
  (b) $5\times10^{-7}$ are shown.
  White dashed lines show $x=0$,
  where the location of the transition between the CZ and the RZ.
  \label{ctbx}}
 \end{figure}

  \begin{figure}[htbp]
  \centering
  \includegraphics[width=16cm]{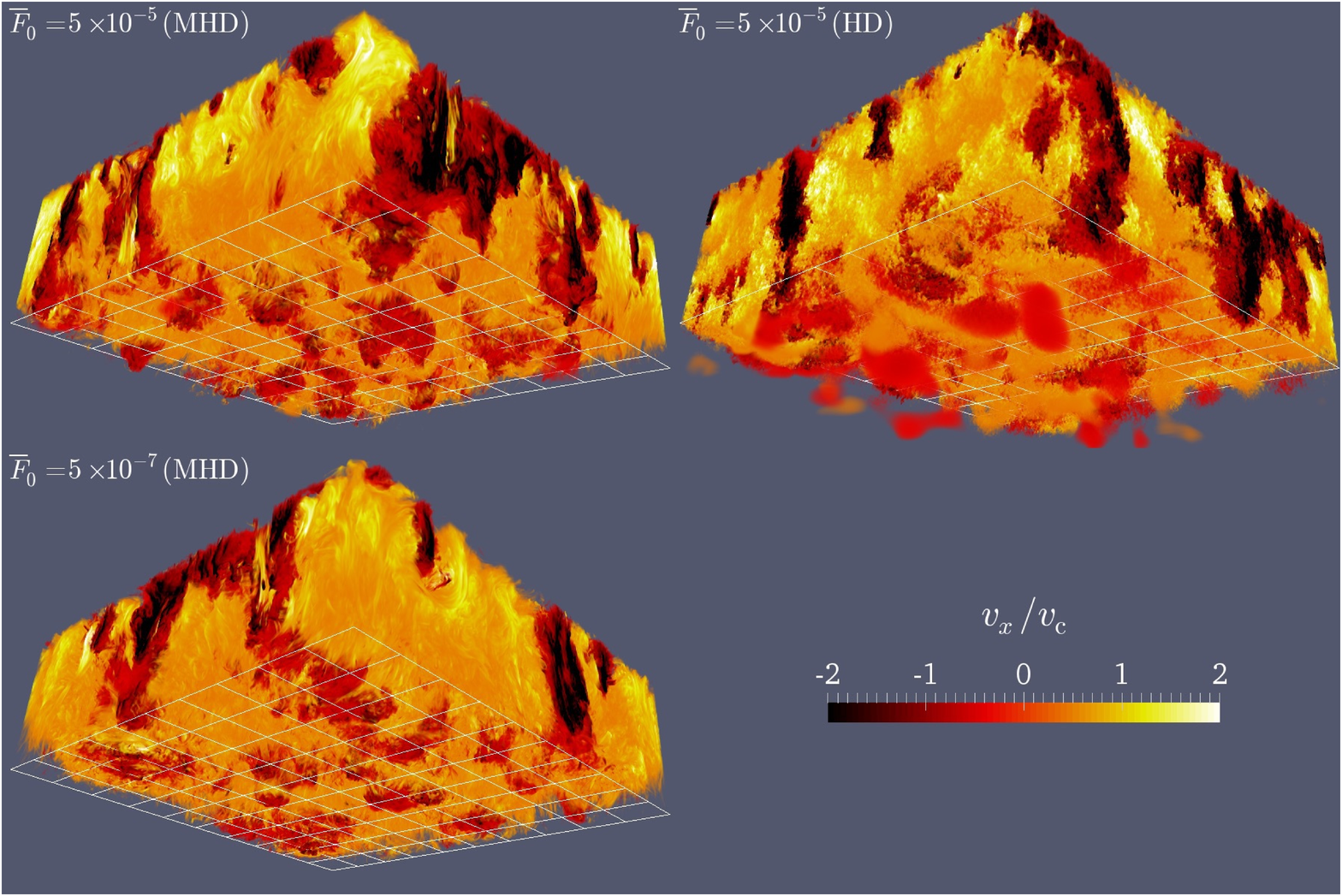}
   \caption{
   3D visualization of the calculations.
   The results with
  $\overline{F}_0=5\times10^{-5}$ with the magnetic field (top-left),
  $\overline{F}_0=5\times10^{-5}$ without the magnetic field (top-right), and
  $\overline{F}_0=5\times10^{-7}$ with the magnetic field (bottom-left)
   are shown.
   Volume rendered
   vertical velocity ($v_x$) is shown. The white grid lines at $x=0$
   show the base of the CZ. Animation is also available online.
  \label{all}}
 \end{figure}

 We now show the overall convection patterns of the calculations.
Fig. \ref{ctvx} shows vertical cuts of the vertical velocity
($v_x$).
The results with the normalized energy
flux of (a) $\overline{F}_0=5\times10^{-5}$ without the magnetic field,
(b) $\overline{F}_0=5\times10^{-5}$ with the magnetic field, and
(c) $\overline{F}_0=5\times10^{-7}$ with the magnetic field are shown.
Perturbations in the RZ ($x<0$) correspond to internal gravity waves.
Since the temperature gradient is almost the same in all cases in the RZ,
the radiative diffusion ($\kappa_\mathrm{r}$) is
proportional to the energy flux imposed at the bottom
boundary. Thus the typical wavelength is short with the small energy
flux (panel c).

As shown in \cite{2015ApJ...803...42H}, an efficient small-scale dynamo
generates strong small-scale magnetic field which significantly suppresses the
small-scale velocity features (Panels a and b).
A comparison between panels
b ($\overline{F}_0=5\times10^{-5}$)
and c ($\overline{F}_0=5\times10^{-7}$) shows that the overall
features of the CZ are similar, but penetration in the
OS is significantly different. Similar features are seen in figures of
the magnetic field.
Figure \ref{ctbx} shows the vertical magnetic field ($B_x$) on a vertical
cut. Panels a and b show the results with
$\overline{F}_0=5\times10^{-5}$ and $5\times10^{-7}$,
respectively. 
Small energy fluxes results in shallow penetration and overall
feature in the CZ ($x>0$) are similar with different energy fluxes.
Three-dimensional visualization nicely shows features of the
calculations in Fig. \ref{all}. An animation of the figure is also
available online.

\par
 
\begin{figure}[htbp]
 \centering
 \includegraphics[width=14cm]{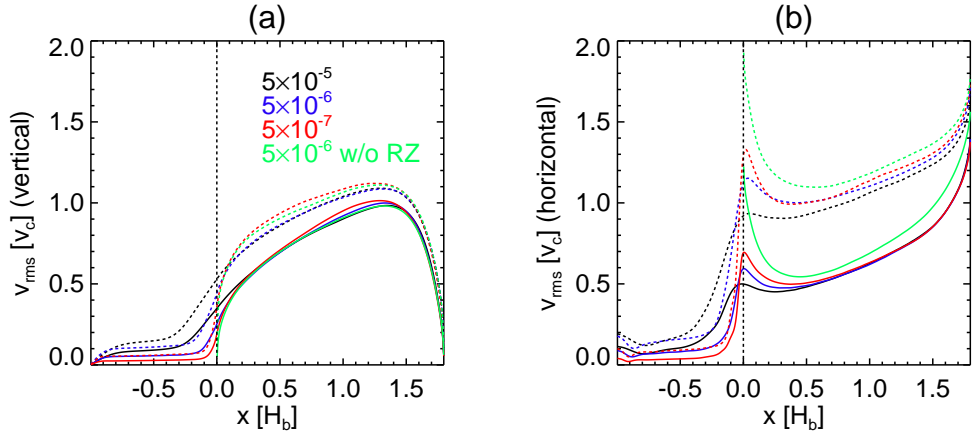}
 \caption{ Vertical ($v_x$: panel a) and horizontal
 ($\sqrt{v_y^2+v_z^2}$: panel b) normalized RMS velocities. 
 Solid and dashed lines show the
 calculations with and without the magnetic field, respectively.
 The results with
 the normalized energy
 flux of $\bar{F}_0=5\times10^{-5}$ (black), $5\times10^{-6}$ (blue), and
 $5\times10^{-7}$(red) are shown. The green lines show the results
 without the RZ and with $\bar{F}_0=5\times10^{-6}$.
 \label{vrms}}
\end{figure}

Figure \ref{vrms} shows the vertical (panel a) and horizontal (panel b)
normalized root-mean-square (RMS) velocities with different
energy fluxes ($\overline{F}_0$).
The RMS velocity is defined as
$v_\mathrm{rms}=\sqrt{\langle (v-\langle v\rangle)^2\rangle}$, where
the bracket indicates the horizontal average.
The result with $\overline{F}_0=5\times10^{-5}$ (black), $5\times10^{-6}$ (blue), and
$5\times10^{-7}$ (red) are shown. Solid and dashed lines show the results
with and without the magnetic field, respectively.
Green lines show the results without the RZ.
As shown in \cite{2015ApJ...803...42H}, the normalized RMS velocity is
significantly suppressed with the magnetic field, due to the high resolution (low diffusivities)
and efficient small-scale dynamo.
As expected, Fig. \ref{ctvx} with a smaller energy flux has a
shallower penetration depth. The details of the penetration depth is
discussed later.
The result without the RZ has the largest horizontal normalized RMS
velocity, since there is artificial
boundary at the bottom. In the cases with the RZ, the smaller energy flux
increases the normalized RMS velocity.
This result indicates that the RZ becomes relatively stiffer with the
smaller energy flux. The circulation of the CZ is then promoted
and a sudden suppression of the downflow at the base of the CZ is observed.
This generates a sudden increase of the horizontal flow at the
base of the CZ (panel b). Interestingly, the results with the magnetic field are
almost the same in cases with different energy flux (solid line). This is caused by the
stronger magnetic field with smaller energy flux. This is also discussed
later.\par

\begin{figure}[htbp]
 \centering
 \includegraphics[width=16cm]{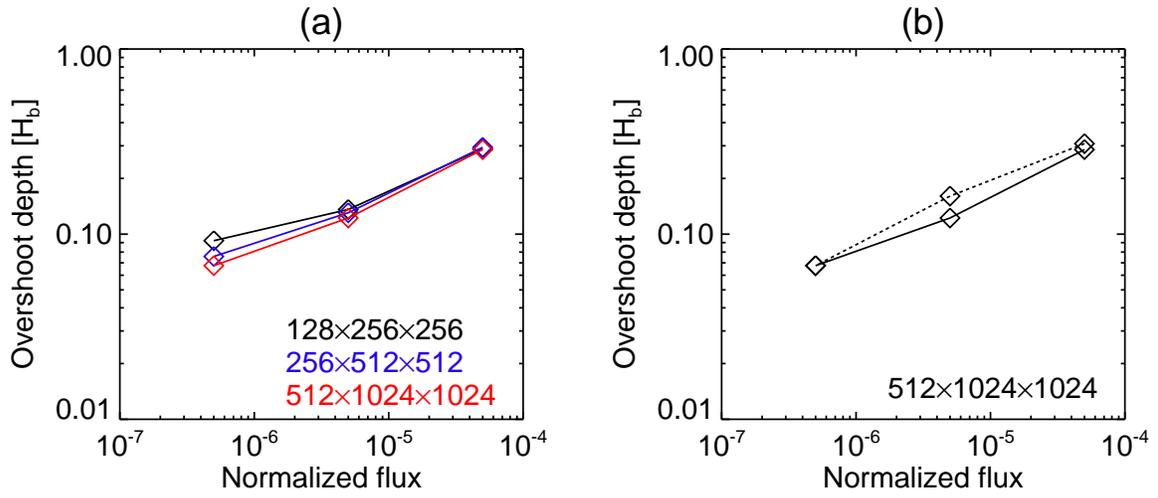}
 \caption{Overshoot depth with different settings is shown. Panel a shows
 dependence of overshoot depth on the energy flux and the resolution.
 The result from the cases Low (black), Medium (blue), and High (red) series
 with the magnetic field are shown.
 Panel b are from the cases High series. Dashed and solid lines show
 the results without and with the magnetic field, respectively.
 \label{penet}}
\end{figure}

Figure \ref{penet} shows the dependence of the overshoot depth
($d_\mathrm{os}$), i.e., the penetration depth dependence on the energy flux and
the resolution. The overshoot depth
is defined with the normalized RMS velocity. We define that the
base of the overshoot region is
the place where the normalized
vertical
RMS velocity is $1/e$ of that at the base of CZ
$x=0$. Panel a shows the results from the cases Low (black),
Medium (blue), and High (red) with the magnetic field.
Smaller energy fluxes make the overshoot depth small.
The cases with different resolutions clearly shows that a higher resolution results
in a shallower penetration.
When we increase the number of grid points, the effect of the artificial
viscosity can be decreased. This makes the overshoot region thin.
The resolution dependence becomes most prominent when the
energy flux is small ($\overline{F}_0=5\times10^{-7}$), since the
overshoot depth is small and a large number of grid points is required.
With the least-square fit with the
results of the cases high resolution series leads reveals a relation between the overshoot
depth and the energy flux as $d_\mathrm{os}/H_\mathrm{b}\sim 6.1\overline{F}_0^{0.31}$. We can
estimate the solar overshoot depth $d_\mathrm{os} \sim0.004H_\mathrm{b}\sim 250\ \mathrm{km}$ when
we extrapolate to $\overline{F_0}\sim5\times10^{-11}$.
Panel b shows a comparison of the results with and without the magnetic field.
The dashed and solid lines shows the
results without and with the magnetic field, respectively.
The cases with the magnetic field tend to show shallower penetrations.
The magnetic field suppresses the convection velocity, significantly in the CZ. Since
the superadiabaticity in the RZ is kept, the RZ becomes relatively
stiff with the magnetic field.
\par

 \begin{figure}[htbp]
 \centering
 \includegraphics[width=16cm]{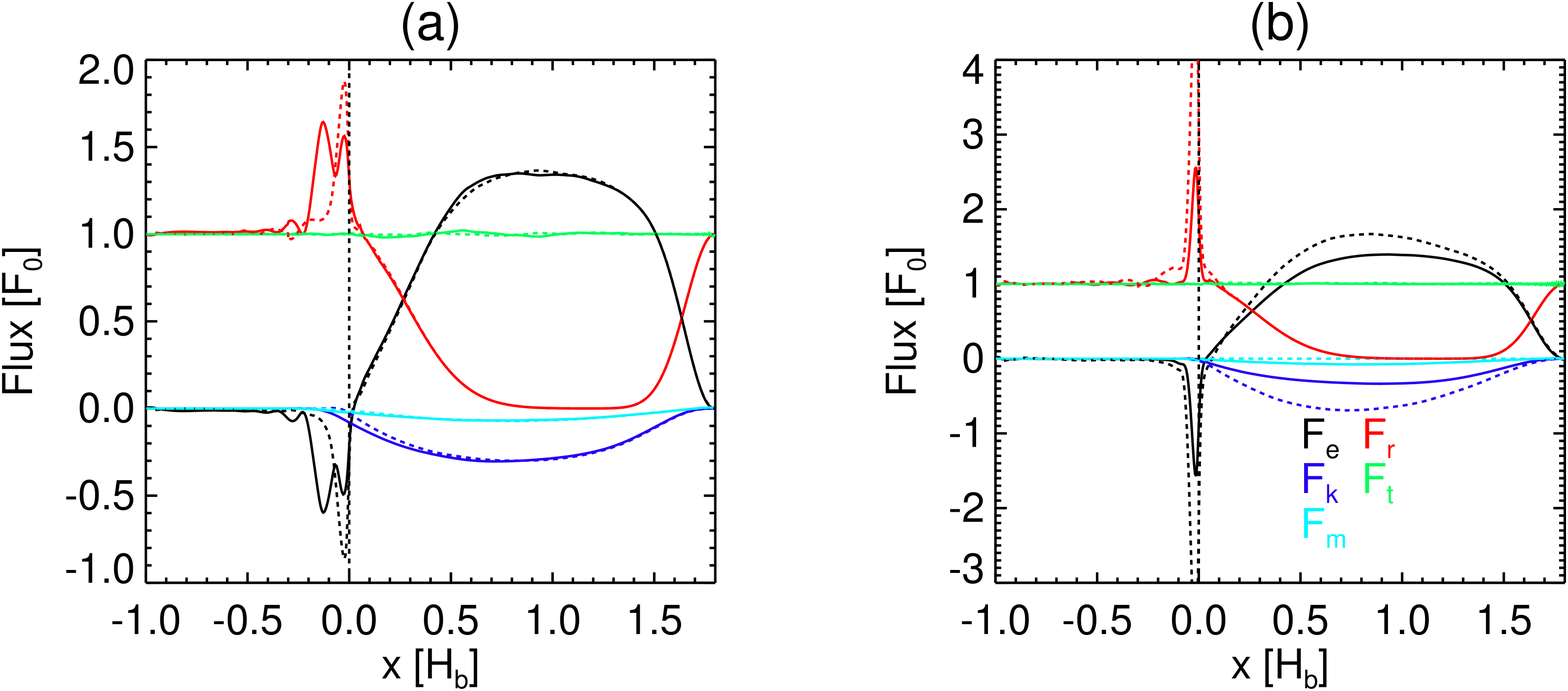}
  \caption{
  The enthalpy ($F_\mathrm{e}$: black), the radiative
  ($F_\mathrm{r}$: red), the kinetic ($F_\mathrm{k}$: blue), the
  Poynting ($F_\mathrm{m}$: light blue) and the total ($F_\mathrm{t}$:
  green) energy flux are shown.
  Here the radiative flux includes the
  artificial cooling in order to avoid complexity in the figure. (a)
  Solid and dashed lines shows the results with
  $\overline{F}_0=5\times10^{-5}$ and $5\times10^{-6}$,
  respectively. Both results have the magnetic field. (b) Solid and
  dashed lines show the result with and  
  without the magnetic field, respectively. The energy flux in both cases is
  $\overline{F}_0=5\times 10^{-7}$.
  \label{flux}}
\end{figure}

Figure \ref{flux} shows the enthalpy ($F_\mathrm{e}$: black), the radiative
($F_\mathrm{r}$: red), the kinetic ($F_\mathrm{k}$: blue), the
Poynting ($F_\mathrm{m}$: light blue) and the total ($F_\mathrm{t}$:
green) energy flux. In order to avoid complexity in the figure, the
radiative flux at the bottom and cooling flux at top boundary (F:
eq. \ref{cooling}) are united
to the radiative flux ($F_\mathrm{r} = -\kappa_\mathrm{r}dT/dr + F$).
Panel a shows the results with $\overline{F}_0=5\times10^{-5}$ (solid) and
$5\times10^{-6}$ (dashed). Both cases include the magnetic
field. In spite of the difference of the energy flux, the overall feature in the CZ is not very much
different. The enthalpy flux in the OS ($x<0$) is much more different in the different cases.
The smaller energy fluxes shows a steep and large negative peak of the enthalpy flux in the OS.
Panel b shows the result with $\overline{F}_0=5\times10^{-7}$ with (solid) and without
(dashed) the magnetic field. The peak of the enthalpy flux is decreased significantly with the magnetic
field in the OS, since the convection velocity is decreased. \par

 \begin{figure}[htbp]
 \centering
 \includegraphics[width=16cm]{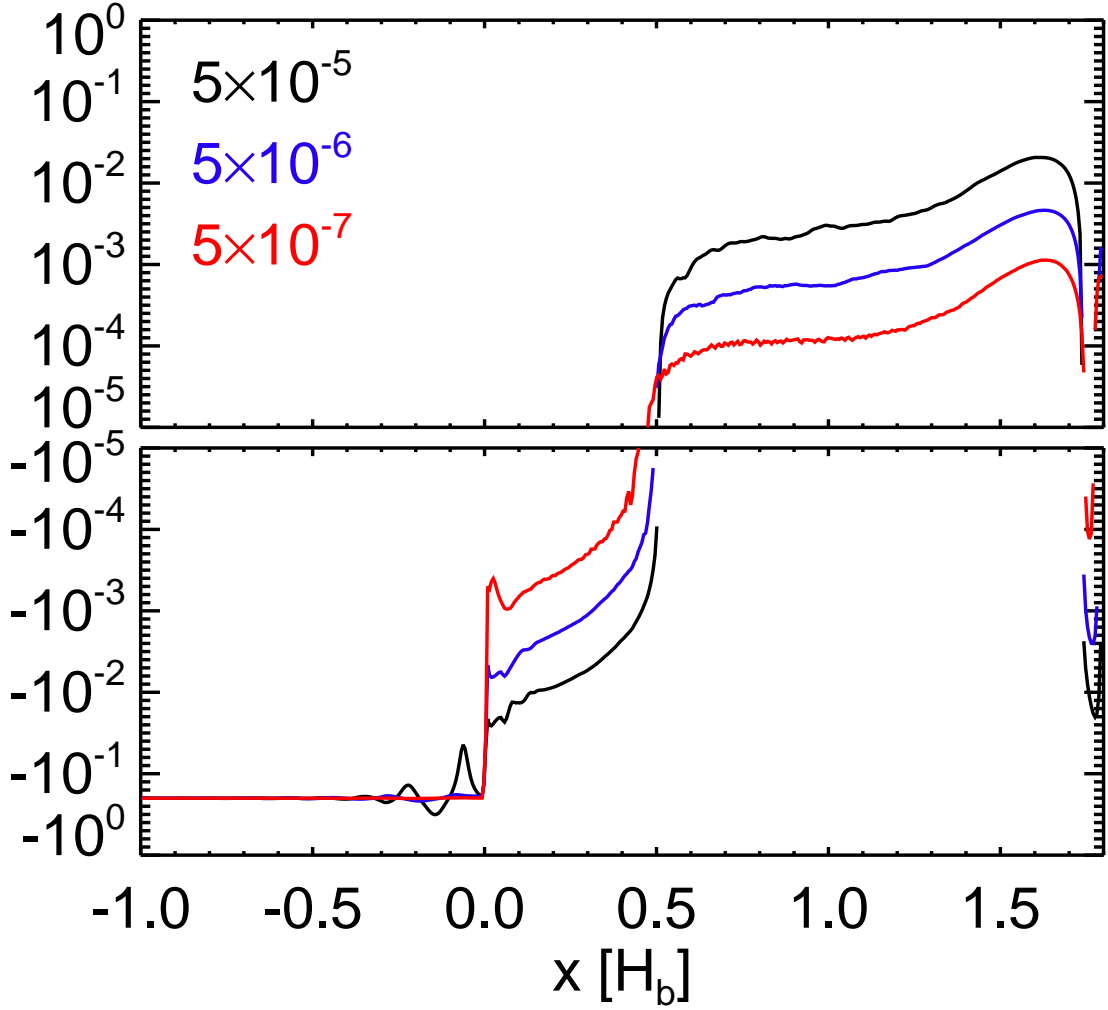}
  \caption{
  Superadiabaticity $\delta$ with different setting is
  shown.
  The results with
  $\overline{F}_0=5\times10^{-5}$ (black), $5\times10^{-6}$ (blue), and
  $5\times10^{-7}$ (red) with the magnetic field are shown.
 \label{strati}}
\end{figure}

Fig. \ref{strati} shows the superadiabaticity.
The results with
$\overline{F}_0=5\times10^{-5}$ (black), $5\times10^{-6}$ (blue), and
$5\times10^{-7}$ (red) with the magnetic field are shown.
As explained in the introduction, the superadiabaticity in the CZ is smaller with
smaller energy flux. An interesting finding is the change of
the sign in the superadiabaticity occurs above the
OS (around $x=0.5H_\mathrm{b}$). This means that a subadiabatic
(convectively stable) region exists in the CZ.
When there is not the subadiabatic layer around the base of
the CZ, the low entropy downflow directly breaks into the RZ. When the
RZ is stiff, most of low entropy fluid is reflected and the low entropy 
accumulates around the base of the CZ. This causes the positive entropy
gradient, i.e., the subadiabatic region around the base of the CZ. This
should exist even in the Sun, because the solar RZ is much stiffer than
what we can achieve in this paper.
The subadiabatic CZ is also found in non-local mixing model
\citep{2004ApJ...607.1046R}.
Once the subadiabatic region is constructed, the downflow is decelerated by the buoyancy even in the CZ.
\par
 \begin{figure}[htbp]
 \centering
 \includegraphics[width=16cm]{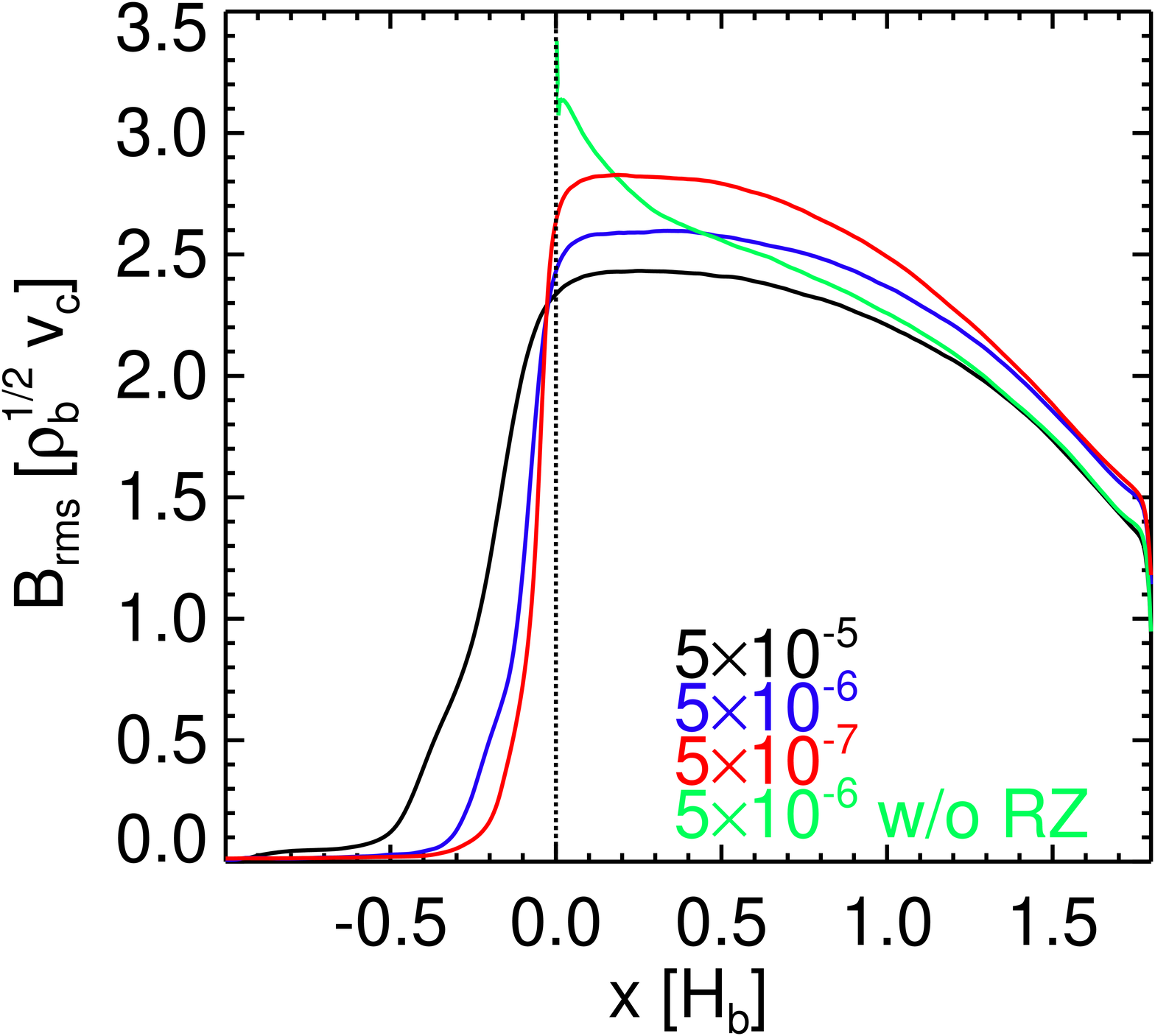}
  \caption{Normalized RMS mangetic field is shown.
  The results with $\overline{F}_0=5\times10^{-5}$ (black),
  $5\times10^{-6}$ (blue), and $5\times10^{-7}$ (red) with
  the RZ are shown. The green line shows the result without the
  RZ. 
 \label{brms}}
 \end{figure}

Figure \ref{brms} shows the normalized RMS magnetic field with different
energy fluxes.
The results with $\overline{F}_0=5\times10^{-5}$ (black),
$5\times10^{-6}$ (blue), and $5\times10^{-7}$ (red) with
the RZ are shown. The green line shows the result without the
RZ. 
As shown in Fig. \ref{ctbx}, shallower penetration with smaller energy
flux is seen. In addition, the normalized RMS magnetic field in the CZ is
increased with a decrease of the energy flux. The case without the RZ is
an extreme of the small energy flux, i.e., stiff RZ.
While the artificial wall increases the magnetic field strength around the base of the CZ,
the magnetic field in the CZ (green) is weaker than in those cases with a RZ.
This could be caused by efficient accumulation of the magnetic field at the
base of the CZ. The result indicates that the efficiently accumulated
magnetic energy at base of the CZ with the artificial boundary is less
transported to the upper part of the CZ than that with the RZ.
Since we cannot achieve the real stiffness of the solar RZ in this study,
the realistic situation should be somewhere between the results with
smallest energy flux and without the RZ. Thanks to the high stiffness of the real sun, the overestimation
introduced by the use of a wall boundary condition is not very severe. \par

  \begin{figure}[htbp]
 \centering
 \includegraphics[width=16cm]{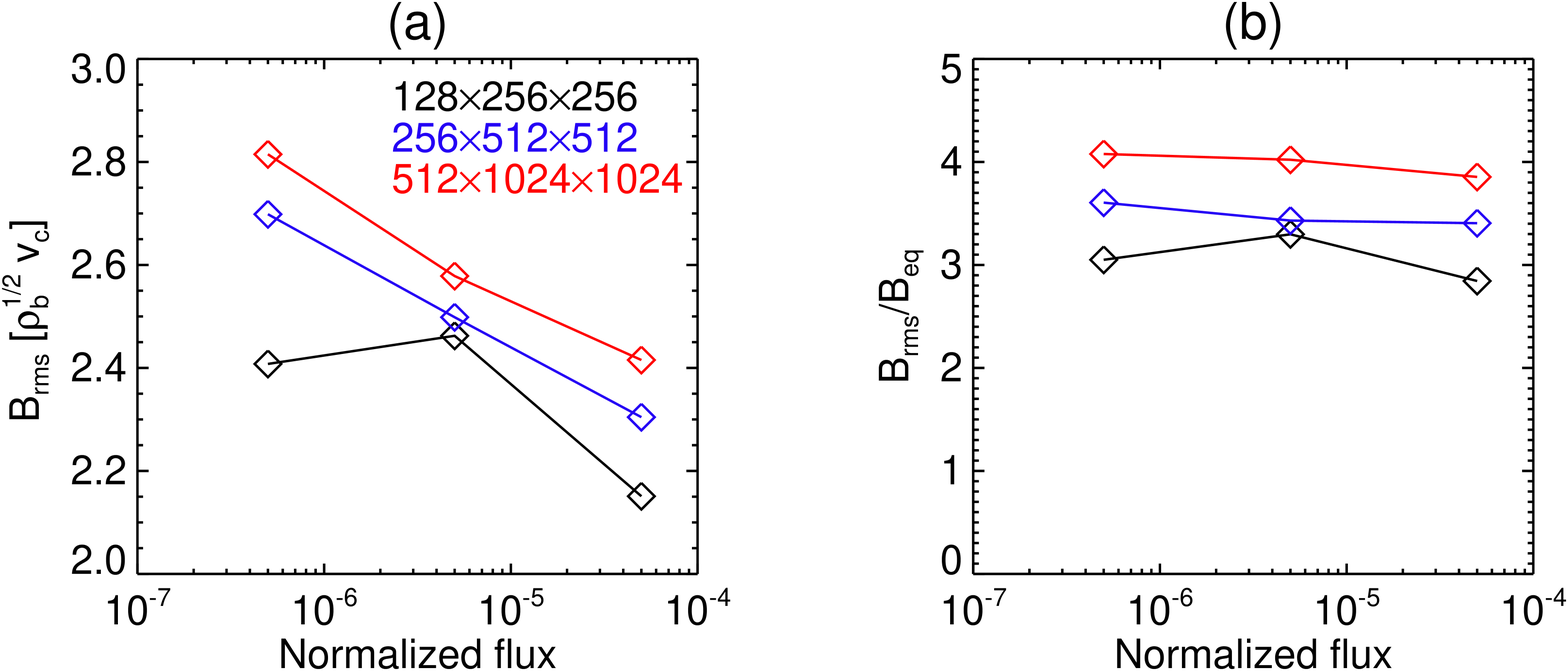}
 \caption{ Dependence of the magnetic field strength on the resolution
   and the energy
   flux is shown.
   The results from
   the case High (black), Medium (blue), and Low series (red) are shown. Panel a and b show
   the normalized RMS magnetic field ($B_\mathrm{rms}$) and the
   normalized RMS
   mangetic field divided by the equipartition magnetic field
    ($B_\mathrm{rms}/B_\mathrm{eq}$) around the base of the CZ ($x=0.1H_\mathrm{b}$),
   respectively.   
 \label{brms_reso}}
 \end{figure}

Figure \ref{brms_reso} shows the dependence of the magnetic field strength
on the resolution and the energy flux. Panel a shows the normalized RMS magnetic
field.
The cases High (black), Medium (blue), and Low (red) series are shown.
The increase of the resolution increases the magnetic field. This is a
general tendency of the small-scale dynamo
\citep{2014ApJ...789..132R,2015ApJ...803...42H}.
The results also indicates that the smaller energy flux tends to produce
stronger magnetic fields. The result with the smallest energy flux
($\overline{F}_0=5\times10^{-7}$) in the cases low resolution series is not
reliable, since the resolution is too low to resolve the thin OS
and the contribution of the artificial viscosity is very large.
Panel b shows the ratio of the normalized RMS magnetic field ($B_\mathrm{rms}$) to
the equipartition magnetic field
($B_\mathrm{eq}=\sqrt{4\pi\rho_0}v_\mathrm{rms}$). This value also
increases with resolution. There is no clear dependence of this
on the energy flux. In the following paragraph, we
investigate the reason why the small energy flux has efficient
small-scale dynamo.
 
   \begin{figure}[htbp]
 \centering
 \includegraphics[width=16cm]{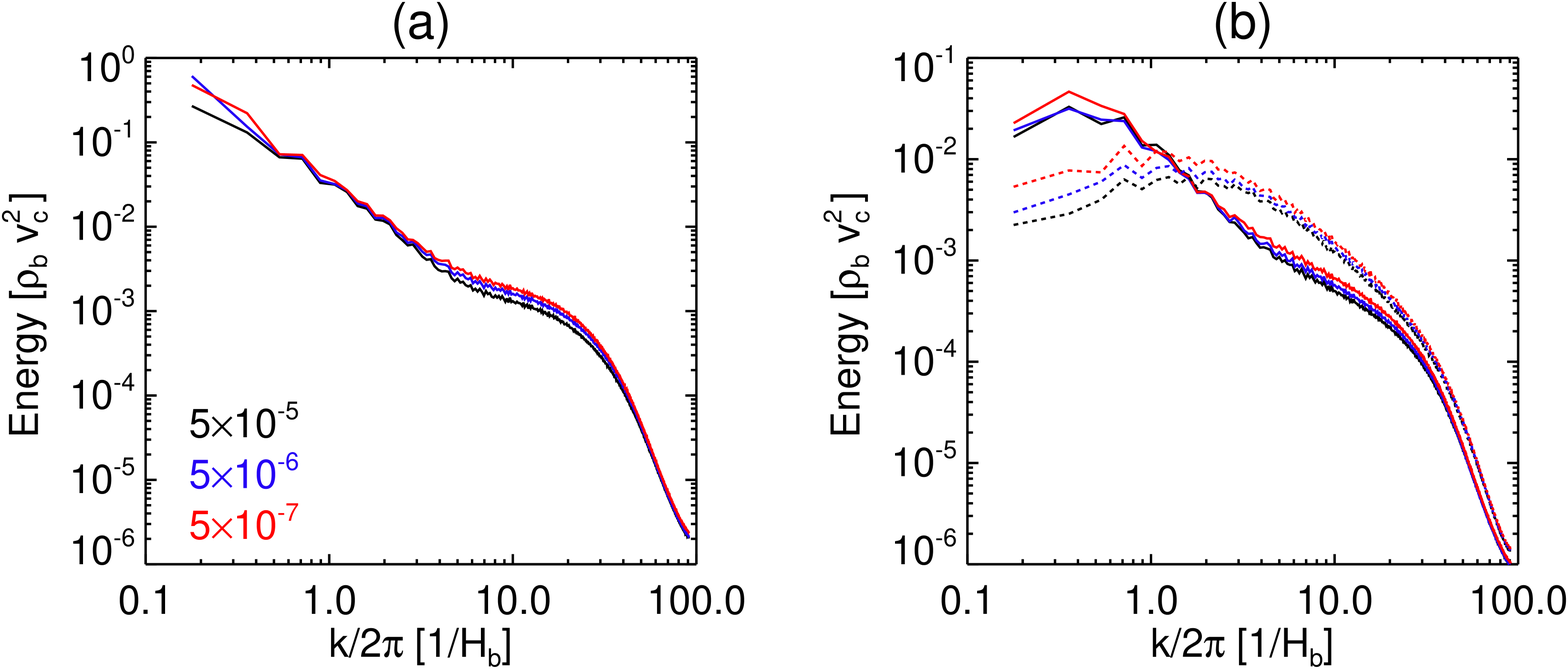}
    \caption{Energy spectra at the base of the CZ ($x=0.1H_\mathrm{b}$) are shown.
    Solid and dashed lines show the kinetic and magnetic energy,
    respectively. Panel a and b show the results without and with the
    magnetic field, respectively.
    The result with $\overline{F}_0=5\times10^{-5}$ (black),
    $5\times10^{-6}$ (blue), and
    $5\times10^{-7}$ (red) are shown.
 \label{spectra}}
 \end{figure}

Figure \ref{spectra} shows the kinetic (solid) and magnetic (dashed)
spectra at the base of the CZ ($x=0.1$). Panel a and b show the results without
and with the magnetic field, respectively
(results with $\overline{F}_0=5\times10^{-5}$ (black),
$5\times10^{-6}$ (blue), and
$5\times10^{-7}$ (red) are shown).
Small energy fluxes have large kinetic energies at small spatial scales
($k/(2\pi) > 1/H_\mathrm{b}$). This could be caused by the stiffer wall
at the base of the CZ. The stiff wall creates the small
scale turbulence with the sudden suppression of the downflow. This
creates the magnetic field through the small-scale dynamo.
\par

  \begin{figure}[htbp]
 \centering
 \includegraphics[width=16cm]{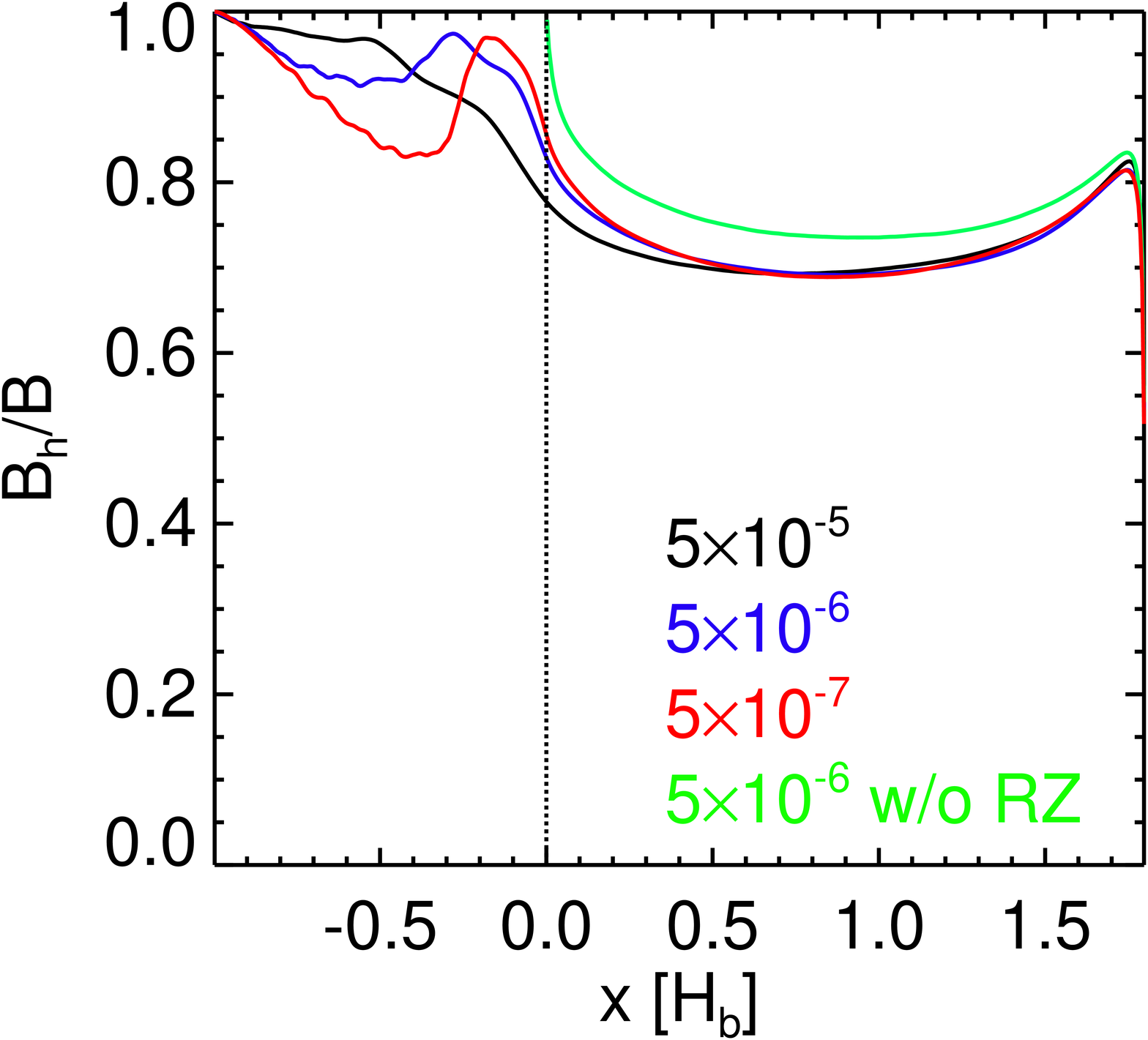}
   \caption{A proxy of the
   inclination of the magnetic field ($B_\mathrm{h}/B$) is
   shown with different   
   energy flux.
   The results with
   $\overline{F}_0=5\times10^{-5}$ (black),
   $5\times10^{-6}$ (blue), and
   $5\times10^{-7}$ (red)with the RZ are shown. The green
   line shows the result with $\overline{F}_0=5\times10^{-6}$ without the RZ.
 \label{inclination}}
  \end{figure}

  Figure \ref{inclination} shows a proxy of the
  inclination of the magnetic field  
  ($B_\mathrm{h}/B$), where $B_\mathrm{h}=(B_y^2+B_z^2)^{1/2}$. The
  mangnetic field becomes horizontal around the base of the CZ when the
  energy flux is small. This is caused by the stiff
  RZ. This is also discussed in the following
  paragraph. Interestingly the inclination in the CZ is not
  influenced by the energy flux. Although the calculation without the RZ
  slightly
  overestimates the inclination with the horizontal magnetic field
  boundary condition, the magnetic field is almost totally horizontal at
  the base of the CZ even with the RZ. We conclude that the horizontal
  magnetic field boundary condition at the base of the CZ is a good
  way to mimic the RZ, when we need to exclude the RZ.

 \begin{figure}[htbp]
 \centering
 \includegraphics[width=16cm]{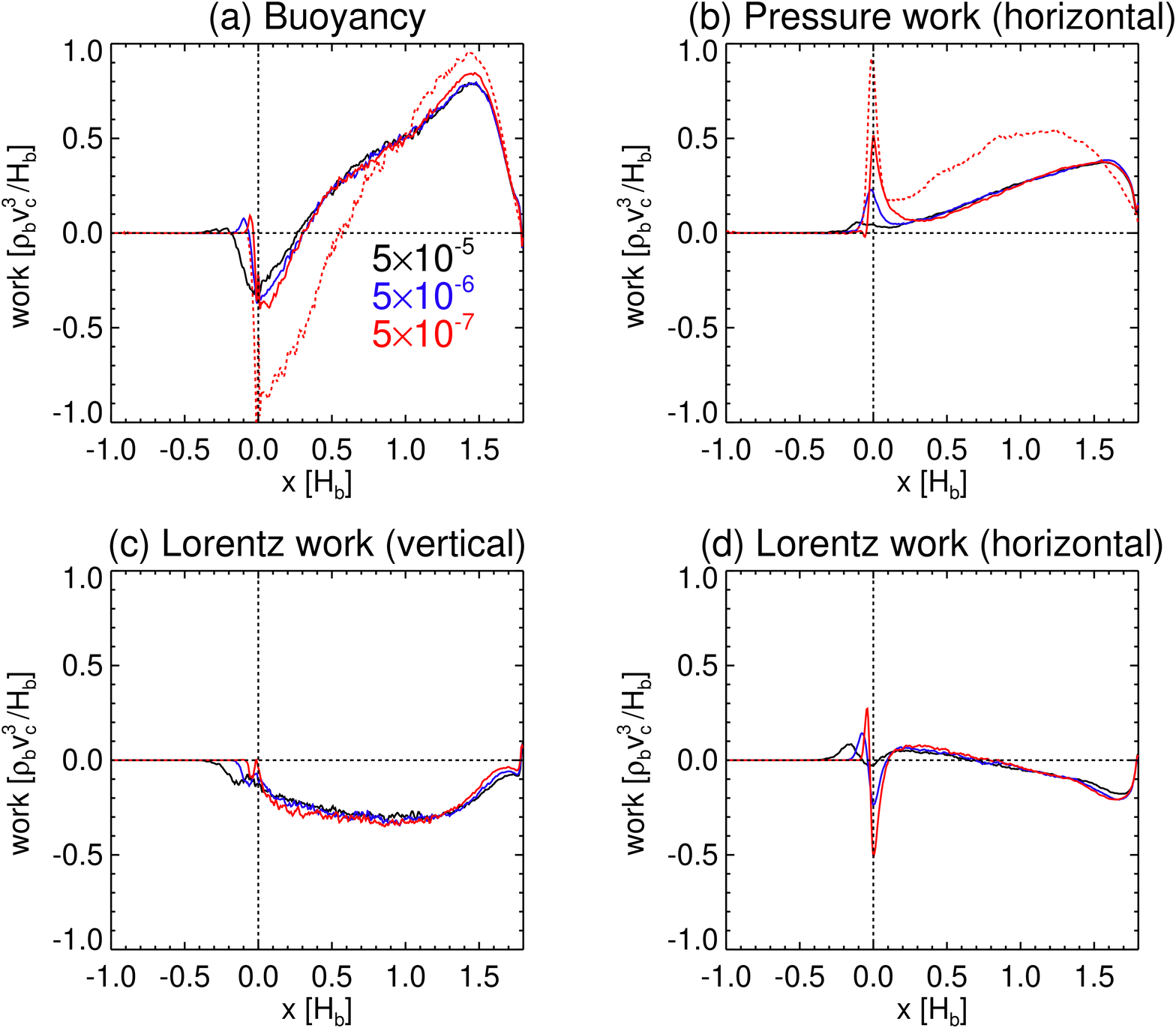}
  \caption{The works are shown with different fluxes.
  The results with
  $\overline{F}_0=5\times10^{-5}$ (black), $5\times10^{-6}$ (blue), and
  $5\times10^{-7}$ (red) with the magnetic field are shown
  The red dashed lines shows the result with
  $\overline{F}_0=5\times10^{-7}$ without the magnetic field.
  The works are averaged over the horizontal plane.
  Each panel shows (a) $W_\mathrm{buo}$, (b)
  $W_\mathrm{pre(h)}$ (c) $W_\mathrm{lor(v)}$, and (d)
  $W_\mathrm{lor(h)}$. The definition of each work is shown in
  eq. (\ref{buo_eq})-(\ref{lorh_eq}).
 \label{works}}
 \end{figure}

In order to investigate the origin of the increase of the magnetic
energy and the field inclination in the cases with the small energy flux,
we analyze energy conversion terms in the equation of motion.
The contributions to the work calculated to this purpose are listed:
 \begin{eqnarray}
  W_\mathrm{buo} &=& -\rho g v_x-\frac{\partial p}{\partial x} v_x, \label{buo_eq}\\
  W_\mathrm{pre(h)} &=& -\frac{\partial p}{\partial y} v_y
   -\frac{\partial p}{\partial z} v_z,\\
  W_\mathrm{lor(v)} &=&
   \frac{\left[{\bf (\nabla\times B)\times B}\right]_x}{4\pi}v_x, \\
  W_\mathrm{lor(h)} &=&
   \frac{\left[{\bf (\nabla\times B)\times B}\right]_y}{4\pi}v_y
   +\frac{\left[{\bf (\nabla\times B)\times B}\right]_z}{4\pi}v_z,
   \label{lorh_eq}
 \end{eqnarray}
where $W_\mathrm{buo}$, $W_\mathrm{pre(h)}$,
$W_\mathrm{lor(v)}$, and $W_\mathrm{lor(h)}$ are the work done by the
buoyancy, the horizontal pressure
gradient, the vertical Lorentz force, and the horizontal Lorentz
force, respectively. Fig. \ref{works} shows horizontally averaged
works with different energy fluxes. 
Panel a shows the buoyancy
work ($W_\mathrm{buo}$).
In the upper CZ which is
superadiabatic, this work is
positive. These works accelerate the thermal convection. Around the
lower CZ, the work becomes negative
due to the subadiabatic CZ. The contribution becomes larger
with decreasing the energy flux. Most of deceleration is done in the CZ
with $\overline{F}_0=5\times10^{-7}$, since the RZ is very stiff and the
OS is very thin.
Although the subadiabtic layer is seen blow $x=0.5H_\mathrm{b}$ in
Fig. \ref{strati}, the deceleration by the buoyancy starts around
$x=0.3H_\mathrm{b}$. This is because the density change in the downflow
does not occur immediately. When the downflow goes into the stable
region from the upper layer, the
downflow is heavier than the mean density. Thus the buoyancy work can be
positive even in the subadiabtic region. After the downflow proceed
certain distance, the density is modified and the buoyancy work becomes negative.
Panel b shows the horizontal pressure
gradient work ($W_\mathrm{pre(h)}$). In most of the CZ, the result
is the same as  with the energy flux, i.e., the work is always positive and
accelerates the thermal convection. On the other hand, the
acceleration at the base of the CZ is significantly depend on the
energy flux. Smaller energy fluxes cause sudden increases of the horizontal
pressure gradient work. When the flow strikes  the stiff
wall, this increases the horizontal pressure gradient and
causes the horizontal flow. Stiff RZ behaves like a wall (we call it ``wall
effect''). Panel c shows the vertical Lorentz force work. The value is
always negative which suppresses the thermal convection. This work does
not depend on the energy flux in the CZ.
Panel d shows the horizontal Lorentz force
work. In the CZ, the work is moderately negative. The most important point
is that the work shows a sudden decrease at the base of the CZ with the small
energy flux. The negative Lorentz work means a gain of the magnetic
energy. The increase of the horizontal velocity at the base of the CZ
due to the wall effect increases the dynamo efficiency. This is the
reason why the small energy flux causes the stronger magnetic energy
at the bottom of the CZ.
The horizontal magnetic field is preferentially amplified and the larger
proxy of the
inclination ($B_\mathrm{h}/B$) is observed.\par
The dashed red lines in panels a and b of Fig. \ref{works} shows the
buoyancy and horizontal pressure works with $F_0=5\times10^{-7}$ without
magnetic field. The significant increase of the buoyancy braking around
the base of the CZ shows the importance of the magnetic field on the
deceleration in the CZ. In magnetic case the horizontal pressure and the
horizontal Lorentz work is almost balanced. In the case without magnetic
field, the horizontal pressure work is significantly increased. This
indicates that the magnetic field suppresses the
horizontal velocity in the CZ.
\section{Comparison with semi-analytic convection/overshoot model}
In order to understand our result properly, it is useful to compare our
result with a semi-analytic convection/overshoot
model. \cite{2004ApJ...607.1046R}
investigates the overshoot feature with several free parameters in broad
range. It is explained that the overshoot characters are mainly determined
with the imposed energy flux $\overline{F}_0$, the filling factor of the
downflow $f$, and the mixing
parameter between up- and downflow $\alpha_\mathrm{d}$.
The energy flux and the filling factor are reduced to a single parameter
$\Phi=\overline{F}_0/f$, which is important to determine the overshoot.
We note that our normalization of the energy flux is a
little different from his expression, which is
$\overline{F}_0=F_0/p_\mathrm{b}(p_\mathrm{b}/\rho_\mathrm{b})^{1/2}$.
When $\Phi$
is large ($>10^{-2}$), the overshoot shows subadiabatic penetration.
On the other hand, small $\Phi$ shows almost adiabatic penetration with
a steep
transition to the RZ. Our calculation shows filling factor of $\sim0.4$
in the
downflow. Thus we estimate $\Phi=1\times10^{-6}$ in our calculation with
the smallest energy flux. Our calculations show almost adiabatic propagation
in the subadiabatic CZ with a steep overshoot of
$d_\mathrm{os}/H_\mathrm{b}\sim6.1\overline{F}_0^{0.31}$. \cite{2004ApJ...607.1046R}
shows similar relation for transition region, in which there is steep
transition from the adiabatic stratification (overshoot region) to the
RZ. The obtained relation in \cite{2004ApJ...607.1046R} is
$d_\mathrm{tr}\sim 5(v_\mathrm{c}/c_\mathrm{s})\sim5\Phi^{1/3}$. Here
the filling factor does not change in our calculations and is the order of
unity. Thus we conclude that the transition region in
\cite{2004ApJ...607.1046R} corresponds to our overshoot
region. Slight change in the power low index between our numerical model
($0.31$) and the semi-analytical model ($1/3$) could be explained with the
insufficient resolution for the thin overshoot region in our
calculation. The thickness of the overshoot region is not numerically
converged and it is expected that higher resolution shows thinner
overshoot with the smallest energy flux (see Fig. \ref{penet}).\par
\cite{2004ApJ...607.1046R} also suggests that the overshoot depth
is mainly determined by the mixing length parameter
($\alpha_\mathrm{d}$), where the mixing length $l$ is expressed as
$l\sim\alpha_\mathrm{d}H_\mathrm{b}$. Smaller $\alpha_\mathrm{d}$ shows
shorter overshoot depth (see Fig. 5 of \cite{2004ApJ...607.1046R}). In
our calculation, we hardly see adiabatic
penetration below the CZ, this indicates a very small
$\alpha_\mathrm{d}$. Strong horizontal flow is caused
around the base of the CZ, this significantly increase the
mixing between up- and downflow in thin layer around the base of the CZ,
this means a small $\alpha_\mathrm{d}$. Fig. \ref{vrms} shows that the
thickness of the strong horizontal flow around the base of the CZ is
about $0.2H_\mathrm{b}$, this efficiently suppresses the downflow in the
CZ. This is seen in the panel a of Fig. \ref{works} which
shows significant suppression of the thermal convection around the base
of the CZ.
In the summary, we conclude that our calculation is consistent with the
model in \cite{2004ApJ...607.1046R} with $\Phi=10^{-6}$ and small
mixing parameter $\alpha_\mathrm{d}$($\sim0.2$). As a result
the adiabatic penetration in the overshoot region is almost disappeared
and the depth of the overshoot region in this paper corresponds to the
transition region in \cite{2004ApJ...607.1046R}.\par
We also note that small $\alpha_\mathrm{d}$ causes small subadiabatic
region in the CZ. In our calculation the subadiabatic
region is extended. This implies $\alpha_\mathrm{d}$ is large in the CZ.
Small $\alpha_\mathrm{d}$ around the base of the CZ is
caused by the sudden suppression of the vertical motion due to the stiff
RZ. Thus $\alpha_\mathrm{d}$ can be dependent on space and larger in the
middle of the CZ, which causes extended subadiabatic region.

 \section{Summary and Discussion}
\begin{figure}[htbp]
  \centering
  \includegraphics[width=16cm]{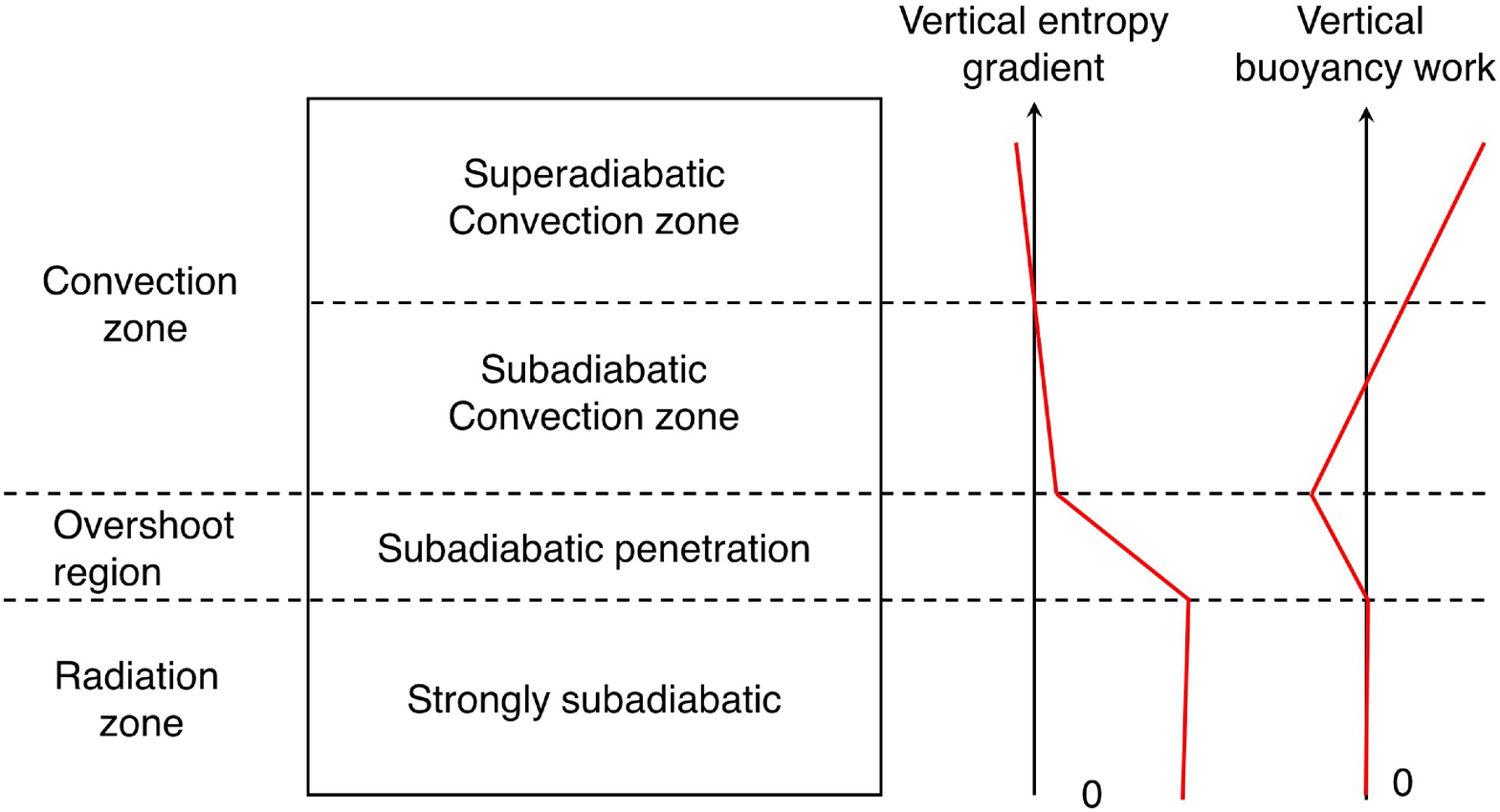}
  \caption{A schematic the CZ, the OS and the RZ describing the vertical
  entropy gradient $ds/dx$ and the vertical buoyancy work
  $-\rho gv_x - \partial p/\partial x v_x$. The quantities are plotted
  on horizontal axis with increasing toward the right.}
  \label{sche}
 \end{figure}

We  have carried out a series of high-resolution calculations of the convection
zone (CZ) and the radiation zone (RZ). The main target of this study is
the interface region between the CZ and the RZ, i.e., the overshoot region (OS).
The difficulty in the investigation of the solar OS is
caused by the small normalized energy flux
($\overline{F_0}=F_0/(\rho_\mathrm{b} c_\mathrm{b}^3)$).
The solar value is $\overline{F}_0\sim5\times10^{-11}$ at the base of
solar CZ. We achieve unprecedentedly small value of the normalized
energy flux ($\overline{F}_0=5\times10^{-7}$) with keeping the
resolution. Most compressible calculations are carried out with
$\overline{F}_0\sim 10^{-3}$ in order to avoid severe constraints of the
sound wave in the CFL condition
\citep[an exception is ][with
$\overline{F}_0\sim10^{-5}$]{2007IAUS..239..437K}. We also investigated
dependence of the small-scale dynamo on the energy flux, i.e., the
stiffness of the RZ.
Our main conclusions are:
\begin{enumerate}
 \item The overshoot depth ($d_\mathrm{os}$) scales with the energy flux
       ($\overline{F}_0$) as
       $d_\mathrm{os}/H_\mathrm{b}=6.1\overline{F}_0^{0.31}$, where
       $H_\mathrm{b}$ is the local pressure scale height. We estimate
       that the real solar overshoot depth is $0.004H_\mathrm{b}\sim
       250\ \mathrm{km}$.
 \item Subadiabatic region is created even in the CZ with similar
       absolute value of the superadiabaticity in the superadiabatic CZ.
 \item Essential deceleration is done in the subadiabatic CZ.
 \item Smaller energy flux, i.e., stiffer RZ makes the small-scale
       dynamo more efficient and the magnetic field more horizontal with sudden
       increase of the horizontal       
       pressure gradient around the base of the CZ.
\end{enumerate}
 Our findings of the structures of the CZ and the RZ are summarized in Fig. \ref{sche}.
 \par
 Although we cannot directly
 address the issue about the large-scale dynamo and magnetic field in
 this study, we can suggest some implications.
 The subadiabatic layer at the bottom half of the CZ would be a good
 place to storage large magnetic flux. The downflow in the top half
 of the CZ efficiently transports the large-scale magnetic flux
 downward. This downflow is decelerated by the subadiabatic layer in the
 CZ and the large-scale magnetic field is less disturbed by the
 downflow.
 \par
 \cite{2002ApJ...570..825B} suggest that the rotation effect decreases
 the depth of the overshoot layer, since the inclination of the downflow
 is changed. It is expected that the rotation works almost the same
 manner even in high resolution and small energy flux achieved in this
 study.
 In addition, the rotation can generates the large-scale
 magnetic field. The large-scale coherent magnetic field efficiently
 suppresses the downflow and decreases the depth of the OS.
 In summary, the rotation effect tends to decrease the depth of the OS
 region. This should be investigated in the future research.
 \par
The absolute value of the superadiabaticity in the subadiabatic CZ
is very small which is similar to that in superadiabatic CZ
($|\delta|\sim \overline{F}_\mathrm{0}^{2/3}$). Using our result, the
expected superadiabaticity around the bottom of the real solar CZ is
$\delta=-1\times10^{-7}$ (subadiabatic).
This is very unlikely  to have an observable influence on the frequency
of the acoustic modes,
i.e., will be very difficult to detect with helioseismology.

Helioseismology has set the upper limit of the overshoot depth of
$0.05H_p$. On the other hand, typical simplified models
underestimate or even ignore the effect of the subadiabatic CZ which has
a main role to suppress the downflow. The overshoot below the CZ is
overestimated ($\sim0.5H_p$). This would be the discrepancy
between the helioseismology and the simplified models.\par
Small density contrast between the bottom and top boundaries
($\rho_\mathrm{b}/\rho_\mathrm{top}\sim 7$)
are used in this study, where $\rho_\mathrm{top}$ is the density at the top boundary. The real
sun has a density contrast of $10^6$.
We note several possible influence of the higher density contrast on the
overshoot. Generally it is expected that if the downflow is coherent
from the solar surface (photosphere) to the bottom of the CZ, the
filling factor of the downflow around the base of CZ becomes very
small. Our previous study, however, with
density contrast of $\sim 600$ also shows the filling factor of 0.4 as
shown
in this study \citep{2014ApJ...786...24H}. This indicates that the
instability around the donwflow
is effective and donwflows cannot be coherent thoughout the entire CZ. Thus
regarding the filling factor, our result could be reliable. Next, the
higher density contrast tends to increase the convection velocity, this
likely increases the overshoot depth. We are possibly underestimating the
overshoot depth a little in this study due to small density contrast.
\par
We note that a recent helioseismic investigation suggests a smooth
transition from the CZ to the RZ
\citep{2011MNRAS.414.1158C}. \cite{2004ApJ...607.1046R} shows that
smooth transition is possible only when the filling factor of the
downflow is tiny and Mach number is large ($\mathrm{Ma}>0.1$) even
around the baser of the CZ or the deceleration is not efficient at the
overshoot region due to low correlation between velocity and
temperature \citep{2001MNRAS.327.1137X}, which is not realized in our
study. The tiny filling factor is achieved only when the downflow is
coherent throughout the CZ with very large density contrast from the
photosphere to the base of the CZ. This would be achievable with an unknown
mechanism to suppress the instability around the downflow.
This is a challenging issue to explore with numerical calculations.
\par
As explained in the introduction, it is difficult to use the anelastic
approximation and the reduced speed of sound technique for
investigations of the overshoot region, since these cannot avoid the high
Brunt-V\"{a}is\"{a}l\"{a} frequency. They are, however,
necessary to cover long time scale large-scale dynamo
calculations. Thanks to the high stiffness of the solar RZ, the
impenetrate boundary with the horizontal magnetic field at the base of the
CZ nicely mimics the solar situation with a slight overestimation of the
magnetic strength and inclination of the magnetic field.
Softer RZ realized with large energy flux or small entropy gradient
should not be used, since this significantly overestimate the storage of
the large-scale magnetic field.

\acknowledgements
The author would like thank anonymous referee for his/her great suggestion which
significantly improve the manuscript.
The author is grateful to M. Rempel, H. Iijima, R. Cameron for their helpful
comments on the manuscript.
The results are obtained by using the K computer at the RIKEN Advanced
Institute for Computational Science (Proposal number hp170239, hp170012, hp160026, hp160252,
ra000008). This work was supported by MEXT/JSPS KAKENHI Grant Number
JP16K17655, JP16H01169.
This research was supported by MEXT as ``Exploratory Challenge on Post-K
computer'' (Elucidation of the Birth of Exoplanets [Second Earth] and
the Environmental Variations of Planets in the Solar System)


\end{document}